\documentclass[twocolumn]{IEEEtran}

\usepackage{amsmath,epsfig,amssymb,verbatim,amsopn,cite,multirow}
\usepackage{amsthm}
\usepackage{balance}
\usepackage{multirow}
\usepackage[usenames,dvipsnames]{color}
\usepackage[all]{xy}  
\usepackage{url}
\usepackage{amsfonts}
\usepackage{amssymb}
\usepackage{epsfig}
\usepackage{epstopdf}
\usepackage{bm}
\usepackage{balance}
\usepackage{graphicx}
\usepackage{subcaption}
\usepackage{footnote}
\usepackage{cancel}
\usepackage{algorithm,algorithmic}

\usepackage[left=1.65cm,right=1.65cm,top=1.8cm,bottom=2.5cm]{geometry}

\newtheorem{Lemma}{Lemma}

\newtheorem{Corollary}[Lemma]{Corollary}

 \usepackage{xcolor}

\newcommand{\qb}{{\bf b}}

\newcommand{\qH}{{\bf H}}
\newcommand{\qI}{{\bf I}}

\newcommand{\qW}{{\bf W}}

\newcommand{\mH}{{\bf{H}}}

\newcommand{\Kh}{\mathcal{K}_h}
\newcommand{\Kl}{\mathcal{K}_l}
\newcommand{\Kw}{\mathcal{K}_w}
\newcommand{\Ks}{\mathcal{K}_s}

\newcommand{\skh}{{\bf s}_{k_h}}

\newcommand{\Htd}{{\bf{H}}^{\rm{TD}}}
\newcommand{\Htf}{{\bf{H}}^{\rm{TF}}}
\newcommand{\Hdd}{{\bf{H}}^{\rm{DD}}}

\newcommand{\FZF}{\mathtt{FZF}}
\newcommand{\PZF}{\mathtt{PZF}}
\newcommand{\MRT}{\mathtt{MRT}}

\newcommand{\BDH}{ \bar{\bf D}^{\rm H}}
\newcommand{\BD}{ \bar{\bf D}}
\newcommand{\D}{ {\bf D}}

\title{How to Combine OTFS and OFDM Modulations in Massive MIMO?}

\author{\IEEEauthorblockN{Ruoxi Chong,
Mohammadali Mohammadi,
Hien Quoc Ngo,
Simon L. Cotton,
and
Michail Matthaiou
}

\IEEEauthorblockA{Centre for Wireless Innovation (CWI), Queen's University Belfast, Belfast, BT3 9DT, United Kingdom\\
Emails: \{rchong02, m.mohammadi, hien.ngo, simon.cotton, m.matthaiou\}@qub.ac.uk}
\vspace{-8mm}
}

\begin{document}

\bstctlcite{IEEEexample:BSTcontrol}

\maketitle

\begin{abstract}
In this paper, we consider a downlink (DL) massive multiple-input multiple-output (MIMO) system, where different users have different mobility profiles. To support this system, we propose to use a hybrid orthogonal time frequency space (OTFS)/orthogonal frequency division multiplexing (OFDM) modulation scheme, where OTFS is applied for high-mobility users and OFDM is used for low-mobility users. Two precoding designs, namely full zero-forcing (FZF) precoding and partial zero-forcing (PZF) precoding, are considered and analyzed in terms of per-user spectral efficiency (SE). With FZF, interference among users is totally eliminated at the cost of high computational complexity, while PZF can be used to provide a trade-off between complexity and performance. To apply PZF precoding, users are grouped into two disjoint groups according to their mobility profile or channel gain. Then, zero-forcing (ZF) is utilized for high-mobility or strong channel gain users to completely cancel the inter-group interference, while maximum ratio transmission (MRT) is applied for low-mobility users or users with weak channel gain. To shed light on the system performance, the SE for high-mobility and low-mobility users with a minimum-mean-square-error (MMSE)-successive interference cancellation (SIC) detector is investigated. Our numerical results reveal that the PZF precoding with channel gain grouping can guarantee a similar quality of service for all users. In addition, with mobility-based grouping, the hybrid OTFS/OFDM modulation outperforms the conventional OFDM modulation for high-mobility users.


\let\thefootnote\relax\footnotetext{This work was supported by a research grant from the Department for the Economy Northern Ireland under the US-Ireland R\&D Partnership Programme. The work of M. Matthaiou was supported  by a research grant from  the European Research Council (ERC) under the European Union’s Horizon 2020 research and innovation programme (grant No. 101001331). The work of H. Q. Ngo was supported by the U.K. Research and Innovation Future Leaders Fellowships under Grant MR/X010635/1.}
\end{abstract}
\vspace{-4mm}
\section{Introduction}
\vspace{-1mm}
Beyond fifth-generation (B5G) wireless communication systems are envisioned to provide reliable communication services under high-mobility environments. In high-mobility scenarios, wireless channels experience doubly dispersive manifestations in the time-frequency (TF) domain. More specifically, time dispersion is caused by the effect of multipath propagation, while frequency dispersion is caused by Doppler shifts. Conventional OFDM systems show great resilience against time dispersion by introducing a cyclic prefix (CP). Unfortunately, the severe Doppler shifts, which appear in high mobility conditions, raise the likelihood of severe inter-carrier interference (ICI). As OFDM systems rely heavily on the orthogonality between different sub-carriers, the performance of currently deployed OFDM systems can be significantly compromised.

In this context, the recently proposed OTFS modulation has been considered as a possible solution to provide reliable communication services under high-mobility conditions, with numerous studies discussing its performance improvement~\cite{Wei2021magzine,Li2021performance,Ruoxi2022achievable ,Ruoxi2022outage}. The advantage of OTFS modulation comes from the introduction of the delay-Doppler (DD) domain signal processing~\cite{Hadani:WCNC:2017,li2022Part2}. The DD domain can be transformed to and from the TF domain using inverse symplectic finite Fourier transform (ISFFT) and SFFT respectively. Thus, under the current OFDM system setup, applying OTFS will entail extra domain transformation processes, resulting in higher computational complexity~\cite{Raviteja:TWC:2018}. In this context, combining OTFS and OFDM, by viewing OTFS as complementary to OFDM, in high-mobility conditions results in a performance-complexity trade-off.

To serve a large number of users, employing different multiple access (MA) schemes for  MIMO-OTFS systems has also become a popular topic~\cite{zhiguoOTFSNOMA,Ruoxi2022SE}.  The coexistence of non-orthogonal MA and OTFS has been investigated in~\cite{zhiguoOTFSNOMA}, while the performance of an orthogonal MA scheme was investigated in~\cite{Ruoxi2022SE}. Moreover, the massive MIMO technology offers the potential for supporting multi-user scenarios with a substantial increase in the SE of OTFS systems~\cite{Wang:JSAC:2020, Shi:TWC:2021, Muye:JSAC:2021, MM2021meet}. More specifically, Liu \emph{et al.}~\cite{Wang:JSAC:2020} and Shen \emph{et al.}~\cite{Shi:TWC:2021} studied OTFS modulation for massive MIMO systems, but only considered channel estimation.  Li \emph{et al.}~\cite{Muye:JSAC:2021} proposed a path division MA scheme for both uplink (UL) and DL transmission for a massive MIMO-OTFS architecture. The achievable DL and UL SEs of cell-free massive MIMO-OTFS systems were analyzed in~\cite{MM2021meet}. However, to the best of our knowledge, the combination of OTFS and OFDM along with different precoding or grouping schemes has not yet been thoroughly studied in the massive MIMO space.

In this paper, we consider a DL massive MIMO system with users from different mobility profiles. We consider a hybrid OTFS/OFDM transmission protocol with different precoding designs. Specifically, we apply OTFS for high-mobility users and OFDM for low-mobility users, while the precoding design type is determined according to the required performance and tolerable complexity. Two different precoding designs are considered at the base station (BS), referred to as FZF and PZF. The former precoding completely suppresses the inter-user interference at the cost of high computational complexity, while the latter balances the complexity and performance at the cost of inter-user interference for some users.  The main technical contributions of this paper can be summarized as follows:
\begin{itemize}
    \item To apply PZF, we consider two user grouping strategies based on two different criteria, 1) mobility profile, referred to as high/low-mobility (HL) user grouping, 2) small-scale channel gain, referred to as strong/weak (SW) user grouping. Then, ZF is applied at the BS  to suppress the inter-group interference at high-mobility or strong user groups, while MRT is used for low-mobility or weak user groups. For each user grouping, ZF and MRT precoding designs are developed, considering hybrid OTFS/OFDM modulation for high/low-mobility users.    
    \item By considering MMSE-SIC detection, we derive new analytical expressions for the DL SE of high- and low-mobility users for different precoding designs as a function of different system parameters.  
    \item Our numerical results reveal that by applying PZF with HL user grouping, high-mobility users achieve significantly higher SE than low-mobility users, while by employing PZF with SW user grouping yields similar SEs for all users.
\end{itemize}
  
\emph{Notations:}
The notations $(\cdot)^{\rm{H}}$ and $(\cdot)^{{\rm T}}$ denote the Hermitian transpose and transpose of a matrix, respectively; ${{{\bf{F}}_N}}$ denotes the normalized discrete Fourier transform (DFT) matrix of size $N\times N$; $\qI_M$ and $\boldsymbol{0}_{M\times N}$ represent the $M\times M$ identity matrix and zero matrix of size $M\times N$, respectively; ``$ \otimes $" denotes the Kronecker product operator; $\textrm{det}(\cdot)$ and $\rm{Tr}(\cdot)$ denote the determinant and trace operations of a matrix, respectively; $\mathbb{E}\{\cdot\}$ denotes the statistical expectation.
\vspace{-1em}
\section{System Model}
Let us consider a DL massive MIMO system consisting of one BS with $N_t$ antennas and $K$ single-antenna users. We consider a general scenario, in which users have heterogeneous mobility profiles e.g., some users are moving at high speeds, denoted by $\Kh \subset \{ 1,\ldots, K \}$, and some users have low-mobility, denoted by $\Kl \subset \{ 1,\ldots, K \}$. Note that $\Kh \cap \Kl = \varnothing$, $K_h=\left| \Kh \right|$, $K_l=\left| \Kl \right|$ and $K_h+ K_l = K$. To achieve a fair balance between complexity and performance, we use the classical OFDM modulation for low-mobility users, whilst OTFS is utilized for high-mobility users. 

At the transmitter side, the information symbols for the ${k_h}$-th high-mobility user $\skh$, ${k_h}\in \Kh$ will first be accommodated onto a DD domain grid of size $M\times N$, where $M$ denotes the number of subcarriers and $N$ is the number of time slots for each frame. Therefore, the DD domain transmitted signal for the $k_h$-th user, ${\bf{x}}_{k_h}^{\rm DD}$, can be expressed as 
\vspace{-0.1em}
\begin{equation}
{\bf{x}}_{k_h}^{\rm DD}={\bf W}_{k_h}  {\bf{s }}_{k_h},
\end{equation}
where ${\bf{s }}_{k_h}$ and ${\bf{x}}_{k_h}^{\rm DD}$ are column vectors of length $MN$ and $N_t MN$, respectively; ${\bf W}_{k_h}$ is the precoding matrix of size $N_tMN\times MN$, and its detailed structure will be exploited later. 

According to the OTFS principle, the time domain transmitted symbol vector for the $k_h$-th user,  ${\bf{x}}_{k_h}^{\rm TD}$, can be obtained by applying the ISFFT and the inverse fast Fourier transform (IFFT)~\cite{Raviteja2019practical},  according to
\vspace{-0.1em}
\begin{align}
		{\bf{x}}_{k_h}^{\rm TD} = \left( {\bf I}_{N_t} \otimes {{{\bf{F}}_N^{\rm H}} \otimes {{\bf{I}}_M}} \right){\bf{x}}_{k_h}^{\rm DD}.
		\label{ntTDTx}
\end{align}

Similarly, for low mobility users ${k_l}$, ${k_l}\in \Kl$, the information symbols ${\bf s}_{k_l}$ will be accommodated and precoded onto an equivalent TF domain grid of size $M\times N$.\footnote{Since a CP is naturally required for OFDM, part of the information symbols for OFDM would be dedicated for CP.} Therefore, we have ${\bf{x}}_{k_l}^{\rm TF}= {\bf W}_{k_l} {\bf{s}}_{k_l}$.
The time domain transmitted symbol vector for the $k_l$-th user,  ${\bf{x}}_{k_h}^{\rm TD}$ can then be obtained by applying IFFT, which is given by ${\bf{x}}_{k_l}^{\rm TD} = \left( {\bf I}_{N_t} \otimes {{{\bf{I}}_N} \otimes {{\bf{F}}^{\rm H}_M}} \right) {\bf{x}}_{k_l}^{\rm TF}$.
Hence, the time domain transmitted signal sent by the BS for each frame can then be denoted by
{
\begin{align}
		{\bf{x}}^{\rm TD} 
		=& \sqrt{E_s} \Big(\sum_{k_h\in \Kh}  \sqrt{\eta _ {k_h}}{\bf{x}}_{k_h}^{\rm TD} +  \sum_{k_l\in \Kl}   \sqrt{\eta _ {k_l}}{\bf{x}}_{k_l}^{\rm TD}\Big),
\end{align}}
where ${E_s}$ is the transmit signal power, while ${\eta _ {k_h}}$ and ${\eta _ {k_l}}$ are the power allocation coefficients for the $k_h$-th and $k_l$-th user, respectively.

We assume that the channel has perfect reciprocity and a total of $P$ independent resolvable paths exist between the BS and each user. Furthermore, we assume that the BS antenna is a uniform linear array with half wavelength inter-element spacing, and define $\phi_{k(i)}$ as the angle of arrival for the $i$-th resolvable path of the $k$-th user.	The steering vector ${\bm \theta} _{k(i)}$ of size $1\times N_t$ can then be denoted by {\small${\bm \theta} _{k(i)}= \big[ 1, {\rm exp}(-j\pi \left(1\right )\sin \phi_{k(i)} ),\ldots , {\rm exp}(-j\pi \left( N_t -1\right )\sin \phi_{k(i)} )\big]$}. With a reduced-CP structure, the time domain channel response between the BS and the $k$-th user can be represented as~\cite{Muye:JSAC:2021}
{\small \begin{align}
\mH_{k}^{\rm TD} 
\!\triangleq \!
 \sum_{i=1}^{P}\! {\bm \theta} _{k(i)} \!\otimes\! {\bf{H}}^{\rm{TD}}_{k (i)}
 \!=\!\!{\sqrt{\beta _{k}}} \!\!\sum_{i=1}^{P}\! {\bm \theta} _{k(i)} \!\!\otimes\! \!\left({h_{k (i)}}  {\bm{\Pi }}^{l_{k (i)}} {\bm{\Delta}}^{k_{k (i)}}\right)\!,
\label{MIMOH}
\end{align} }
where $\beta _{k}$ is the large-scale fading coefficient for the $k$-th user; $h_{k(i)}$ is the small-scale fading coefficient of the $i$-th path, which follows the Gaussian distribution with zero mean and $1/(2P)$ variance per real dimension; ${\bf{\Pi }}$ is a permutation matrix (forward cyclic shift) of size $MN\times MN$ characterizing the delay effect, i.e., $\boldsymbol{\Pi} =\mathrm{circ}\{[0,1,0,\ldots,0]^{{\rm T}}_{MN\times 1}\}$, and ${\bm{\Delta}}=\textrm{diag}\{{\alpha}^0,{\alpha}^1,\ldots,{\alpha}^{MN-1}\} $ is a diagonal matrix characterizing the Doppler effect with ${\alpha} \buildrel \Delta \over = {e^{\frac{{j2\pi }}{{MN}}}}$~\cite{Raviteja2019practical}. Furthermore, the terms  $l_{k (i)}$ and $k_{k (i)}$ in~\eqref{MIMOH} are the indices of delay and Doppler associated to the $i$-th path, respectively.\footnote{Note that \eqref{MIMOH} gives a tight approximation when the system has fractional Doppler indices~\cite{ViterboBook}.}  

Therefore, the received signal in the time domain for the $k$-th user can be denoted by
{
\begin{align}
		{\bf y}^{\rm TD}_{(k)} 
		&=  \Htd _{k} {\bf{x}}_{k}^{\rm TD}+\sum_{k'=1, k'\neq k}^{K} \Htd _{k} {\bf{x}}_{k'}^{\rm TD}+ {{\bf z}_{k}},
		\label{inout}
\end{align}}
 where ${{\bf z}_{k}}$ is the additive white Gaussian noise (AWGN) sample vector, with ${\mathbb E}\left\{ {{\bf z}_{k}} {{\bf z}_{k}}^{\rm H} \right\}={\bf I}_{ MN}$.  Without loss of generality, we assume that ${{\beta_k \eta_k}}=1$, which corresponds to the inverse power control scheme \cite{blandino2017link}.
We now define the DD domain and TF domain equivalent channel matrices as follows
\begin{align}
\mH_{k}^{\rm DD}
=& ({{{\bf{F}}_N} \otimes {{\bf{I}}_M}}) \Htd _{k} \left( {{\bf I}_{N_t} \otimes{{\bf{F}}_N^{\rm H}} \otimes {{\bf{I}}_M}} \right),\\
\mH_{k}^{\rm TF}
=&({ {{\bf{I}}_N} \otimes  {{\bf{F}}_M} }) \Htd _{k} \left( {{\bf I}_{N_t} \otimes{{\bf{I}}_N} \otimes {{\bf{F}}^{\rm H}_M}} \right).
\end{align}
The equivalent DD domain received signal for the $k_h$-th high-mobility user, and the equivalent TF domain received signal for the $k_l$-th low-mobility user are shown in~\eqref{inout_kh} and~\eqref{inout_kl}, respectively, at the top of the next page.
\begin{figure*}
{\small
\begin{align}
{\bf y}^{\rm DD}_{(k_h)} &= ({{{\bf{F}}_N} \otimes {{\bf{I}}_M}}) {\bf y}^{\rm TD}_{(k_h)} 
=\underbrace{  \sqrt{E_s}   \Hdd _{k_h}   {\bf W}_{k_h} {\bf{s }}_{k_h} }_{\text{Desired signal}}
+\!\!\underbrace{\!\! \sum_{k'_h\in \Kh,\atop k'_h\neq k_h} \!\!\!\! \sqrt{E_s}   \Hdd _{k_h} {\bf W}_{k'_h}  {\bf{s }}_{k'_h}}_{\text{Intra-group interference}} \! 
+ \underbrace{ \!\!\!\!\sum_{k_l\in \Kl} \!\! \sqrt{E_s}  ({{{\bf{F}}_N} \otimes {{\bf{I}}_M}}) \Htd _{k_h} \left( {{\bf I}_{N_t} \otimes{{\bf{I}}_N} \otimes {{\bf{F}}^{\rm H}_M}} \right)  {\bf W}_{k_l}  {\bf{s }}_{k_l} }_{\text{Inter-group interference}} +{{\bf z}_{k_h}},
		\label{inout_kh}
\end{align}
}
{\small
\begin{align}
{\bf y}^{\rm TF}_{(k_l)} 
&= ({ {{\bf{I}}_N} \otimes  {{\bf{F}}_M} }) {\bf y}^{\rm TD}_{(k_l)} 
=\underbrace{\sqrt{E_s}   \Htf _{k_l} {\bf W}_{k_l} {\bf{s }}_{k_l} }_{\text{Desired signal}}
+\underbrace{\!\!\!\!\sum_{k'_l\in \Kl,\atop k'_l\neq k_l}\!\! \sqrt{E_s}   \Htf _{k_l}   {\bf W}_{k'_l} {\bf{s }}_{k_l'} }_{\text{Intra-group interference}} 
+\!\! \underbrace{\sum_{k_h\in \Kh}\!\! \sqrt{E_s} ({ {{\bf{I}}_N} \otimes  {{\bf{F}}_M} }) \Htd _{k_l} \left( {{\bf I}_{N_t} \otimes{{{\bf{F}}_N^{\rm H}} \otimes {{\bf{I}}_M}}} \right)   {\bf W}_{k_h}  {\bf{s }}_{k_h} }_{\text{Inter-group interference}} +{{\bf z}_{k_l}}.
\label{inout_kl}
\end{align}
}
\centering
\rule{\textwidth}{0.3mm}
\vspace{-3em}
\end{figure*}
For notation simplicity, we further define 
\begin{align}
{{\Bar\mH}_{k}^{\rm DD}}
=& ({{{\bf{F}}_N} \otimes {{\bf{I}}_M}}) \Htd _{k} \left( {{\bf I}_{N_t} \otimes{{\bf{I}}_N} \otimes {{\bf{F}}^{\rm H}_M}} \right),\\
{\Bar\mH_{k}}^{\rm TF}
=&({ {{\bf{I}}_N} \otimes  {{\bf{F}}_M} }) \Htd _{k} \left( {{\bf I}_{N_t} \otimes{{{\bf{F}}_N^{\rm H}} \otimes {{\bf{I}}_M}}} \right).
\end{align}
\vspace{-6mm}
\section{Performance Analysis}
We consider an MMSE-SIC detector at the receiver side and analyze the SE performance of different precoding designs for different user grouping strategies.  The DL achievable SE can be obtained as~\cite{SE}
\vspace{-1mm}
\begin{align}~\label{eq:SEk}
{\rm{SE}}_k
&=\alpha_{\rm{SE}} {\log _2}\det \left( {\bf{I}}_{MN} +  \BDH_{k k} \left( \Psi_k \right)^{-1} \BD_{k k} \right),
\end{align}
where ${\small \BD_{k k}=\mathbb{E}\left\{ {\D}_{k k} \right\}}$, and 
\begin{subequations}
{\small
	\begin{align}
		&\D_{k k} = \sqrt{{E_s}}  \qH_{k}   {\bf W}_{k} ,\\
		&\Psi_k = {\bf I}_{MN} + \mathbb E \Big\{\sum^{K}_{k'=1} \D_{k k'} \D_{k k'}^{\rm H}  \Big\} - \BD_{k k} \BDH_{k k} \label{21c},
	\end{align}  
 }
\end{subequations}
with $\D_{k k'} = \sqrt{E_s}  \qH_{k}   {\bf W}_{k'}$, and $\qH_{k}$ represents the MIMO channel for the $k$-th user. The specific $\qH_{k}$ for each case (i.e., $\mH_{k}^{\rm DD}$, ${{\Bar\mH}_{k}^{\rm DD}}$,  $\mH_{k}^{\rm TF}$, and ${{\Bar\mH}_{k}^{\rm TF}}$) will be made explicit later according to the user grouping strategy and modulation scheme. Moreover, $\alpha_{\rm{SE}}$ is a normalization coefficient, where $\alpha_{\rm{SE}}=\alpha_{\rm{SE}}^{\rm DD}=\frac{1}{MN}$ for OTFS users, and $\alpha_{\rm{SE}}=\alpha_{\rm{SE}}^{\rm TF}=\frac{1}{MN} \frac{MN-L_{cp}}{MN}$ for OFDM users. We notice that~\eqref{eq:SEk} results in a tight approximation to the real system SE due to the channel hardening effect provided by the massive MIMO system.
\vspace{-0,7em}
\subsection{FZF precoding}
\vspace{-0.3em}
Let us first consider FZF for all users. 
With the grouping method based on users' mobility, we further define 
{\footnotesize${\mH}^{\rm FZF}_{\rm H}=[(\Hdd _{1})^{\rm T}\!,\!(\Hdd _{2})^{\rm T}\!,\!\ldots\!,\!(\Hdd_{K_h})^{\rm T}\!,\!({\Bar\mH_{1}}^{\rm TF})^{\rm T}\!,\!({\Bar\mH_{2}}^{\rm TF})^{\rm T}\!,\!\ldots\!,\!({\Bar\mH_{K_l}}^{\rm TF})^{\rm T}]^{\rm T}$}, and 
{\footnotesize${\mH}^{\rm FZF}_{\rm L}\!\!=\!\![(\Htf _{1})^{\rm T}\!,\!(\Htf _{2})^{\rm T}\!,\!\ldots\!,\!(\Htf_{K_l})^{\rm T}\!,\!({\Bar\mH_{1}}^{\rm DD})^{\rm T}\!,\!({\Bar\mH_{2}}^{\rm DD})^{\rm T}\!,\!\ldots\!,\!({\Bar\mH_{K_h}}^{\rm DD})^{\rm T}]^{\rm T}$}. 
Note that the sizes of ${\mH}^{\rm FZF}_{\rm H}$ and ${\mH}^{\rm FZF}_{\rm L}$ are both $KMN\times N_tMN$. Besides, we have $  \big ( ({{\bf{b}}_{ K }^{(k_h)}})^{\rm H} \otimes {\bf I}_{MN} \big){\mH}^{\rm FZF}_{\rm H} = \Hdd _{k_h}$, where ${{\bf{b}}_{ K }^{(k_h)}}$ is a column vector of length $K$, with only the $k_h$-th entry being one and others being zero.  Therefore, we can see that $\big ( ({{\bf{b}}_{ K }^{(k_h)}})^{\rm H} \otimes {\bf I}_{MN} \big)$ helps to pick out the $k_h$-th matrix from the block matrix ${\mH}^{\rm FZF}_{\rm H}$. 
Then, for the high-mobility user, we have
\vspace{-0,2em}
\begin{equation}
\qW_{k_h}^\FZF\!=\!\alpha_{\FZF,{\rm H}} 
		({\mH}^{\rm FZF}_{\rm H})^{\rm H} \!\!\left (  {\mH}^{\rm FZF}_{\rm H} ({\mH}^{\rm FZF}_{\rm H})^{\rm H} \right )^{\!\!-1}\!\! \big ( {\qb_{K}^{(k_h)}} \!\otimes  \qI_{MN} \!\big) ,
  \label{FZF}
\end{equation}
where $\small{\alpha_{\FZF,{\rm H}} = \frac{{\sqrt{MN}}}{\sqrt{\mathbb E \left\{ \left\| ({\mH}^{\rm FZF}_{\rm H})^{\rm H} \left (  {\mH}^{\rm FZF}_{\rm H} ({\mH}^{\rm FZF}_{\rm H})^{\rm H}\right )^{-1} \left ( {{\bf{b}}_{K}^{(k_h)}} \otimes {\bf I}_{MN} \right) \right\| ^2 \right\} }}}$ is the normalization coefficient, with
\vspace{-0.1em}
{\small
\begin{align}
&{{\mathbb E \Big\{ \big\| ({\mH}^{\rm FZF}_{\rm H})^{\rm H} \left (  {\mH}^{\rm FZF}_{\rm H} ({\mH}^{\rm FZF}_{\rm H})^{\rm H}\right )^{-1} \big ( {{\bf{b}}_{K}^{(k_h)}} \otimes {\bf I}_{MN} \big) \big\| ^2 \Big\} }}\notag\\
&=\mathbb E \Big\{{\rm Tr} \Big[\left ( ({{\bf{b}}_{K}^{(k_h)}})^{\rm H} \otimes {\bf I}_{MN} \right)  \left (  {\mH}^{\rm FZF}_{\rm H} ({\mH}^{\rm FZF}_{\rm H})^{\rm H} \right )^{-1} {\mH}^{\rm FZF}_{\rm H}\notag \\
   &~~({\mH}^{\rm FZF}_{\rm H})^{\rm H} \left ( {\mH}^{\rm FZF}_{\rm H}({\mH}^{\rm FZF}_{\rm H})^{\rm H} \right )^{-1} \left ( {{\bf{b}}_{K}^{(k_h)}} \otimes {\bf I}_{MN} \right)  \Big] \Big\} \notag\\
&=\frac{1}{K}{{\mathbb E \left\{  {\rm Tr} \left[  \left (  {\mH}^{\rm FZF}_{\rm H}({\mH}^{\rm FZF}_{\rm H})^{\rm H}  \right )^{-1}    \right ] \right\} }}.
\label{alpha_FZF}
\end{align}}
The normalization coefficient $\alpha_{\FZF,{\rm H}}$ can be further expressed as 
{\small
\begin{align}
\alpha_{\FZF,{\rm H}}
=\frac{{\sqrt{K MN}}}{\sqrt{{\mathbb E \left\{ {\rm Tr} \left[  \left (  {\mH}^{\rm FZF}_{\rm H}({\mH}^{\rm FZF}_{\rm H})^{\rm H}  \right )^{-1}    \right ] \right\} }}}.
\end{align}
}
Similarly, for the $k_l$-th low-mobility user, we have  
{\small
\begin{equation}
\qW_{k_l}^\FZF\!=\!\alpha_{\FZF,{\rm L}} 
		({\mH}^{\rm FZF}_{\rm L})^{\rm H} \!\!\left (  {\mH}^{\rm FZF}_{\rm L} ({\mH}^{\rm FZF}_{\rm L})^{\rm H} \right )^{\!\!-1}\!\! \left ( \!{\qb_{K}^{(k_l)}} \!\otimes \!\qI_{MN} \!\right) ,
\end{equation}}
with $\alpha_{\FZF,{\rm L}}
=\frac{{\sqrt{KMN}}}{\sqrt{{\mathbb E \left\{  {\rm Tr} \left[  \left (  {\mH}^{\rm FZF}_{\rm L}({\mH}^{\rm FZF}_{\rm L})^{\rm H}  \right )^{-1}    \right ] \right\} }}}$. Given the Independence of the different users' channel responses and the unitary characteristics of the domain transformation, we have $\alpha_{\FZF} \buildrel \Delta \over=\alpha_{\FZF,{\rm H}}=\alpha_{\FZF,{\rm L}}$.

\noindent\textbf{Proposition 1}:~\label{theorem:FZF:LM}
With FZF, the SE for the $k_h$-th high-mobility and the $k_l$-th low-mobility user can be obtained in closed-form as
\vspace{-0.8em}
\begin{subequations}
	\begin{align}
		{\rm{SE}}_{k_h}^\FZF&= MN\alpha_{\rm{SE}}^{\rm DD} {\log _2} \left( 1 +  \alpha_{\FZF}^{2} {{E_s}}  \right), \label{SE_HFZF} \\
		{\rm{SE}}_{k_l}^\FZF&=MN\alpha_{\rm{SE}}^{\rm TF} {\log _2} \left( 1 +  \alpha_{\FZF}^{2} {{E_s}}  \right).
     \label{SE_LFZF}
	\end{align}
 \end{subequations}
\begin{proof}
    See Appendix~\ref{APX:theorem:FZF:LM}.
\end{proof}

From Proposition 1, we observe that with the help of the FZF, all the intra-group interference and inter-group interference can be canceled at the cost of high computational complexity for both high-mobility and low-mobility users. 

\vspace{-1em}
\subsection{PZF Precoding with High/Low-mobility Grouping}
Enabling FZF requires high complexity, alternatively, we now consider a precoding design with lower complexity. Specifically, we consider ZF precoding for high-mobility users and MRT precoding for low-mobility users. 

For the high-mobility users, the precoding matrix $\qW_{k_h}^\PZF$ can be denoted by
\begin{equation}
		\qW_{k_h}^\PZF\!=\!\alpha_{\PZF} 
		(\mH^\PZF)^{\rm H} \left (  \mH^\PZF (\mH^\PZF)^{\rm H} \right )^{\!\!-1} \big ( {\qb_{K_h}^{(k_h)}} \otimes \qI_{MN} \!\big) ,
  \label{PZF}
\end{equation}
where $\mH^\PZF=[(\Hdd _{1})^{\rm T},(\Hdd _{2})^{\rm T},\ldots,(\Hdd_{K_h})^{\rm T}]^{\rm T}$ of size $K_hMN\times N_tMN$, and
{\small
\begin{align}
\alpha_{\PZF}= \frac{{\sqrt{MN}}}{\sqrt{\mathbb E \Big\{ \big\| (\mH^\PZF)^{\rm H} \left ( \mH^\PZF (\mH^\PZF)^{\rm H}\right )^{-1} \big ( {{\bf{b}}_{K_h}^{(k_h)}} \otimes {\bf I}_{MN} \big) \big\| ^2 \Big\} }},
\end{align}  }
is the normalization coefficient.  
Therefore, for PZF, we have,
{\small
\begin{align}
&\mathbb E \Big\{ \big\| (\mH^\PZF)^{\rm H} \left ( \mH^\PZF (\mH^\PZF)^{\rm H}\right )^{-1} \left ( {{\bf{b}}_{K_h}^{(k_h)}} \otimes {\bf I}_{MN} \right) \big\| ^2 \Big\}\notag\\
&=\mathbb E \Big\{{\rm Tr} \Big[\left ( ({{\bf{b}}_{K_h}^{(k_h)}})^{\rm H} \otimes {\bf I}_{MN} \right)  \left (  {\mH}^{\PZF} ({\mH}^{\PZF})^{\rm H} \right )^{-1} {\mH}^{\PZF}\notag \\
&\hspace{2em}\times({\mH}^{\PZF})^{\rm H} \left ( {\mH}^{\PZF} ({\mH}^{\PZF})^{\rm H} \right )^{-1} \left ( {{\bf{b}}_{K_h}^{(k_h)}} \otimes {\bf I}_{MN} \right)  \Big] \Big\} \notag\\
&= \frac{1}{K_h}{{\mathbb E \left\{ {\rm Tr} \left[  \left (  {\mH}^{\PZF}({\mH}^{\PZF})^{\rm H}  \right )^{-1}    \right ] \right\} }},
\end{align}}
and the normalization coefficient can be denoted by
\begin{align}
\alpha_{\PZF} 
=\frac{\sqrt{MNK_h}}{\sqrt{{{\mathbb E \left\{ 
{\rm Tr} \left[  \left (  {\mH}^{\PZF}({\mH}^{\PZF})^{\rm H}  \right )^{-1}    \right ] \right\} }}}}.
\end{align}
From~\eqref{FZF} and~\eqref{PZF}, we can see that the main complexity for computing the FZF and PZF precoding designs can be approximated by $O((KMN)^3)$ and $O((K_hMN)^3)$, respectively.
Therefore, the complexities of FZF and PZF are dependent on $K$ and $K_h$, respectively, with PZF having lower complexity.

For low-mobility users, to apply the MRT precoding, we have 
\begin{equation}
		\qW_{k_l}^\MRT=\alpha_{\MRT}{(\Htf _{k_l})}^{\rm H},
  \label{W_MRT}
\end{equation}
where $\alpha_{\MRT}=\frac{\sqrt{MN}}{\sqrt{\mathbb E \big\{ \big\| \Htf _{k_l} \big\| ^2 \big\} }}$, while
{\small
\begin{align}
	&\mathbb E \big\{ \big\| \Htf _{k_l} \big\| ^2 \big\} 
 =\mathbb E\left\{{\rm Tr}\left [ \Htf _{k_l} (\Htf  _{k_l})^{\rm H} \right ]\right\}\notag\\
&=\mathbb E\Big\{{\rm Tr} \Big[({ {{\bf{I}}_N} \otimes ( {{\bf{F}}_M^{\rm H}}{{\bf{F}}_M}) }) \Htd _{k_l} \left( { {{\bf{I}}_{N_t N}}} \otimes ({{\bf{F}}_M^{\rm H}}{\bf{F}}_M)  \right)  \left ( \Htd_{k_l }\right )^{\rm H}  \Big]\Big\}\notag \\
&=\mathbb E\bigg\{\sum_{i=1}^{P}{\rm Tr}\left [ \left ({\bm \theta} _{k_l(i)} {\bm \theta}^{\rm H} _{k_l(i)} \right ) \otimes    \left (  {\bf{H}}^{\rm{TD}}_{k_l(i)}  ({\bf{H}}^{\rm{TD}}_{k_l (i)})^{\rm H} \right )\right ]\bigg\}\notag \\
&~~+\mathbb E\bigg\{\!\sum_{i=1}^{P}\!\!\sum_{j=1,j\neq i}^{P}\!\!{\rm Tr}\left [ \left ({\bm \theta} _{k_l(i)} {\bm \theta}^{\rm H} _{k_l(j)} \right ) \otimes \left({\bf{H}}^{\rm{TD}}_{k_l(i)} ({\bf{H}}^{\rm{TD}}_{k_l (j)})^{\rm H} \right )\right ]\bigg\}\notag \\
&\stackrel{(a)}=\sum_{i=1}^{P}\mathbb E\left\{ {\rm Tr}\left [  {{h_{k_l (i)}}} {{h_{k_l (i)}^{*}}} N_t {\bf I}_{MN}\right ] \right\}
=N_tMN,
	\label{a_n_mrt}
\end{align}
}
where $(a)$ in~\eqref{a_n_mrt} follows the fact that the zero-mean channel coefficients for different paths are independent from each other. Therefore, the normalization coefficient becomes
\vspace{-0.2em}
\begin{align}
\alpha_{\MRT} 
=\frac{1}{\sqrt{N_t}}.
\label{alpha_MRT}
\end{align}
\textbf{Proposition 2}:~\label{theorem:PZFMRT}
 With PZF precoding and HL grouping, for the $k_h$-th high-mobility user, the SE can be represented as 
 \begin{align}
{\rm{SE}}_{k_h}^\PZF\!&=\!\alpha_{\rm{SE}}^{\rm DD} {\log _2}\!\det \!\bigg( {\bf I}_{MN}\!+\! E_s \alpha_\PZF^2 \big ( 1 \!+\! \frac{K_l {E_s} \alpha_\MRT ^2}{P^2} \notag\\
&~~\times \sum_{i=1}^{P} \sum_{j=1}^{P} 
\mathbb E \left\{ | {\bm \theta} _{k_h(i)} {\bm \theta} _{k_l'(j)}^{\rm H} |^2 \right\} 
\big )^{-1} {\bf I}_{MN} \bigg).
 \label{SE_PZF_MRT_h}
\end{align} 
\begin{proof}
See Appendix~\ref{Ptheorem:PZFMRT}.    
\end{proof}
\vspace{-3mm}

\noindent\textbf{Proposition 3}:~\label{theom:PZFMRT}
With PZF precoding and HL grouping,  the SE for the $k_l$-th low-mobility user is given by
{\small
\begin{align}
{\rm{SE}}_{k_l}^\MRT
&=\alpha_{\rm{SE}}^{\rm TF} {\log _2}\det \Big(
{\bf I}_{MN}+E_s N_t
\!\Big( 
{\bf I}_{MN} 
- E_s N_t {\bf I}_{MN}
 \notag\\
& + K_h  \mathbb E \big\{ \D_{k_l k_h'} \!\D_{k_l k_h'}^{\rm H}  \big\}+\mathbb E \big\{ \D_{k_l k_l'} \D_{k_l k_l'}^{\rm H}  \big\} 
\nonumber\\
&+ \frac{{E_s} \alpha_\MRT ^2}{P^2} \sum_{k_l'\neq k_l}\sum_{i=1}^{P} \sum_{j=1}^{P} \mathbb E \left\{ | {\bm \theta} _{k_l(i)} {\bm \theta} _{k_l'(j)}^{\rm H} |^2 \right\}  {\bf I}_{MN} 
\Big)^{-1}
\Big),
 \label{SE_PZF_MRT_l}
\end{align}
}
\vspace{-1mm}
where 
{\small
\begin{align}
\mathbb E \big\{ \D_{k_l k_h'} \D_{k_l k_h'}^{\rm H}  \big\}
&={E_s} \alpha_{\PZF}^2 \mathbb E \big\{{\Bar{\qH}}^{\rm TF}_{k_l} 
 (\mH^\PZF)^{\rm H} \left (  \mH^\PZF (\mH^\PZF)^{\rm H} \right )^{\!-1} \notag\\
&\hspace{-6.5em}\times\!\left ( \!{\qb_{K_h}^{(k_h')}} ({\qb_{K_h}^{(k_h')}})^{\rm H}\! \otimes \! \qI_{MN} \!\right) 
\left (  \mH^\PZF (\mH^\PZF)^{\rm H} \!\right )\!^{-\! 1}\mH^\PZF
({\Bar{\qH}}^{\rm TF}_{k_l})^{\rm H}\big\},
\label{Dkkp_2}
\end{align}  }
\vspace{-1mm}
{\small
\begin{align}
\mathbb E \big\{ \D_{k_l k_l'} \D_{k_l k_l'}^{\rm H}  \big\} 
&={E_s} \alpha_\MRT ^2 \bigg ( 
 \frac{P+1}{P} N_t^2\notag \\
&\hspace{1em}+\frac{1}{P^2} \sum_{i=1}^{P} \sum_{j=1 \atop j \neq i}^{P}  \mathbb E \left\{ | {\bm \theta} _{k_l(i)} {\bm \theta} _{k_l(j)}^{\rm H} |^2 \right\} \bigg )
{\bf I}_{MN}.
\label{DD_klkl}
\end{align} } 
\vspace{-1.5mm}
\begin{proof}
\vspace{-2mm}
By invoking~\eqref{W_MRT} and~\eqref{alpha_MRT}, for the $k_l$-th low mobility user, we have
{\small
\begin{align}
\BD_{k_l k_l} 
&=  \sqrt{E_s} \mathbb{E}\left\{  {\qH}^{\rm TF}_{k_l}   {\bf W}_{k_l}^\MRT \right\}  
={\sqrt {N_t E_s}} {\bf I}_{MN}.
\end{align}  }
Then, similar to~\eqref{Dkkp_1}, for the intra-group interference from user $k_l'$, 
where $k_l'\in \Kl$, $k_l'\neq k_l$, we have
\begin{align}
\mathbb E \big\{ \D_{k_l k_l'} \D_{k_l k_l'}^{\rm H}  \big\} &={E_s} \mathbb E \left\{{{\qH}}^{\rm TF}_{k_l} {\bf W}_{k'_l}^\MRT ({\bf W}_{k'_l}^\MRT)^{\rm H} ({{\qH}}^{\rm TF}_{k_l})^{\rm H}\right\} \notag\\
&\hspace{-2em}= \frac{{E_s} \alpha_\MRT ^2}{P^2} 
\sum_{i=1}^{P} \sum_{j=1}^{P} 
\mathbb E \left\{ | {\bm \theta} _{k_l(i)} {\bm \theta} _{k_l'(j)}^{\rm H} |^2 \right\} 
{\bf I}_{MN}. 
\label{Dklklp}
\end{align} 
Similar as in~\eqref{Dklklp}, with $k_l'= k_l$, $\mathbb E \big\{ \D_{k_l k_l} \D_{k_l k_l}^{\rm H}  \big\} $ is expressed as in~\eqref{DD_klkl}, which is derived based on the independency between different paths. 
To this end, after computing $\Psi_k$ according to~\eqref{21c} and then plugging the result into~\eqref{eq:SEk} we arrive at the final result in~\eqref{SE_PZF_MRT_l}. 
\end{proof}
\vspace{-1em}

\begin{Corollary}
The achievable SE with PZF precoding and HL grouping in~\eqref{SE_PZF_MRT_h} and~\eqref{SE_PZF_MRT_l} can be further approximated as
\begin{align}
{\rm{SE}}_{k_h}^\PZF&\approx  {\log _2} \Big( 1+\! E_s \alpha_\PZF^2 \Big ( 1 \!+\! {K_l {E_s} \alpha_\MRT ^2} N_t
\Big )^{-1}  \Big).
 \label{SE_PZF_MRT_h1}\\
{\rm{SE}}_{k_l}^\MRT&\approx \alpha_{\rm{SE}}^{\rm TF} {\log _2}\det \bigg({\bf I}_{MN}+E_s N_t \nonumber\\
&\hspace{2em} \times\Big ( \! K_h  \mathbb E \big\{ \D_{k_l k_h'} \D_{k_l k_h'}^{\rm H} \!\big\}+ 
\psi
{\bf {I}}_{MN}  \Big )\!^{-1}  \bigg),
 \label{SE_PZF_MRT_l1}
\end{align} 
with $\psi\!=\! 1 \!+\! \left ( N_t\!\!+\!\frac{N_t\!-\!1}{P}  \!+\! K_l\right ){E_s} \alpha_\MRT ^2 N_t \!-\! E_s N_t $. 
Note that the approximation comes from $\mathbb E \left\{{\bm \theta} _{k_l'(j)}^{\rm H} {\bm \theta} _{k_l'(j)}\right\} \approx {\bf I}_{MN} $.
\end{Corollary}

According to Propositions 2 and 3, with the help of the PZF, intra-group interference for high-mobility users can be eliminated. Yet, high-mobility users still suffer from inter-group interference, while low-mobility users will experience both intra-group and inter-group interference.
 \subsection{PZF Precoding with Strong/Weak User Grouping}
From~\eqref{SE_PZF_MRT_h} and~\eqref{SE_PZF_MRT_l}, we can see that under the HL user grouping strategy, low-mobility users always experience more interference than high-mobility users. To guarantee a satisfactory performance for both high-mobility and low-mobility users, based on~\eqref{inout_kh} and~\eqref{inout_kl}, we consider further grouping the users based on their channel gain as strong users $\Ks \subset \{ 1,\ldots, K \}$ and weak users $\Kw \subset \{ 1,\ldots, K \}$, with $\Ks \cap \Kw = \varnothing$, $K_s=\left| \Ks \right|$, $K_w=\left| \Kw \right|$ and $K_s+ K_w = K$.  Therefore, each user's modulation protocol is decided by its mobility, and its corresponding precoding design is determined based on its channel conditions.
Based on the SW grouping strategy,  we apply  PZF precoding for strong users and MRT precoding for  weak users.

Without loss of generality, we again focus on one high-mobility and one low-mobility user. With the SW user grouping, the input-output relationship can be denoted as in~\eqref{inout_sw_s} and~\eqref{inout_sw_w} at the top of the next page.
\begin{figure*}
{\small
\begin{align}
{\bf y}^{\rm DD}_{(k_h)} 
\!= \!\!\!\! \sum_{k'_h\in \Kh \cup \Ks}\!\!\!\! \!\sqrt{E_s}   \Hdd _{k_h}   {\bf W}_{k_h'}^\PZF {\bf{s }}_{k_h'} 
\!+\!\!\!\sum_{k'_h\in \Kh \cup \Kw}\!\!\!\!\! \!\sqrt{E_s}   \Hdd _{k_h}   {\bf W}_{k_h'}^\MRT {\bf{s }}_{k_h'} 
\!+\!\!\!\sum_{k'_l\in \Kl \cup \Ks}\!\!\!\!\!\! \!\sqrt{E_s}   {{\Bar\mH}_{k_h}^{\rm DD}}   {\bf W}_{k_l'}^\PZF {\bf{s }}_{k_l'} 
\!+\!\!\!\sum_{k'_l\in \Kl \cup \Kw}\!\!\!\!\!\! \sqrt{E_s}   {{\Bar\mH}_{k_h}^{\rm DD}}   {\bf W}_{k_l'}^\MRT {\bf{s }}_{k_l'} \!+\!{{\bf z}_{k_h}},
\label{inout_sw_s}
\end{align}
}
\vspace{-1mm}
{\small
\begin{align}
{\bf y}^{\rm TF}_{(k_l)} 
\!= \!\!\!\! \sum_{k'_h\in \Kh \cup \Ks}\!\!\!\! \sqrt{E_s}   {\Bar{\mH}_{k_l}^{\rm TF}}   {\bf W}_{k_h'}^\PZF {\bf{s }}_{k_h'} 
\!+\!\!\!\sum_{k'_h\in \Kh \cup \Kw}\!\!\!\! \sqrt{E_s}   {\Bar{\mH}_{k_l}^{\rm TF}}  {\bf W}_{k_h'}^\MRT {\bf{s }}_{k_h'} 
\!+\!\!\!\sum_{k'_l\in \Kl \cup \Ks}\!\!\!\! \sqrt{E_s}   {{\mH}_{k_l}^{\rm TF}}   {\bf W}_{k_l'}^\PZF {\bf{s }}_{k_l'} 
\!+\!\!\!\sum_{k'_l\in \Kl \cup \Kw}\!\!\!\! \sqrt{E_s}   {{\mH}_{k_l}^{\rm TF}}   {\bf W}_{k_l'}^\MRT {\bf{s }}_{k_l'} \!+\!{{\bf z}_{k_l}}.
\label{inout_sw_w}
\end{align}
}
\centering
\rule{\textwidth}{0.3mm}
\vspace{-3.4em}
\end{figure*}
From~\eqref{inout_sw_s}, we can see that, if the $k_h$-th user is also a strong user, the desired signal is in the first term when $k_h'=k_h$, or if the $k_h$-th user is a weak user, the desired signal is in the second term when $k_h'=k_h$. Similar to before, with PZF, the intra-group interference from the weak users to the strong users can be eliminated. However, strong users still experience inter-group interference from weak users. Weak users are affected by both intra-group and inter-group interference. With the help of the SW grouping strategy, as each user can be a strong or a weak user with equal probability, the performance gap between the high-mobility and low-mobility users can be narrowed. Due to space limitations, this result will only be verified with simulations in Section~\ref{sec:num}.

\vspace{-2mm}
\section{Numerical Results}~\label{sec:num}
We now simulate the per-user SE with different modulation and precoding designs, adopting the Monte Carlo approach. We set $N_t=100$, $M=8$, $N=8$, and $P=2$. The channel fading coefficient is generated with a uniform power delay profile. Similar to~\cite{Li2021performance,li2021cross}, the  delay $l_{k(i)}$ and Doppler indices $k_{k (i)}$ are generated with equal probability within the
range of $[0, l_{max}]$ and $[-k_{max}, k_{max}]$, where the maximum delay index $l_{max}\!=\!4$ and the maximum Doppler index $k_{max}\!=\!2$ for low-mobility users, $l_{max}\!=\!4$ and $k_{max}\!=\!4$ for high-mobility users, respectively. Note that in the simulations,  we consider non-fractional delay indices and fractional Doppler indices, which resemble practical channel conditions. Besides, we consider $20\%$ of the overall transmitted symbols as CP for OFDM.

\begin{figure}
\centering
\includegraphics[width=0.43\textwidth]{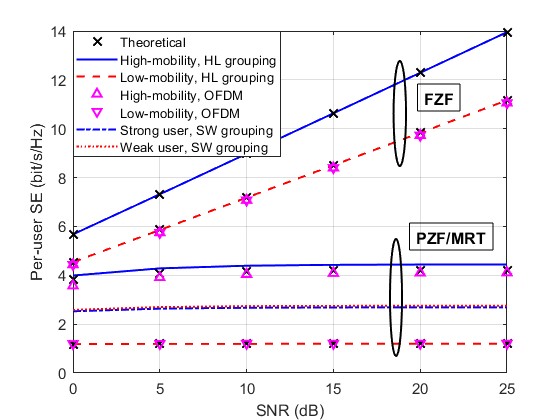}
\caption{Per-user SEs for FZF and PZF precoding with different modulation schemes ( $K_h=K_l=K_s=K_w=3$).}
\label{TF_DD}
\vspace{0.5 em}
\centering
\end{figure}

Figure~\ref{TF_DD} shows the simulation results for per-user SE achieved by FZF and PZF precoding designs with the hybrid OTFS/OFDM transmission under the HL and SW user grouping strategies and for $K_h=K_l=K_s=K_w=3$. From this figure, it can be seen that the performance enhancement of FZF over PZF in terms of SE, especially when the SNR increases. Moreover, with PZF precoding design and HL user grouping, low-mobility users experience a worse overall performance as they suffer from more interference. 
On the other hand, for the SW user grouping, it can be observed that the SE performance is similar for both high- and low-mobility users. Therefore, users who experience a weak channel gain, are protected at the cost of the strong users' performance.
Moreover, the results validate our theoretical analysis. Note that the theoretical values in the figure are calculated via~\eqref{SE_PZF_MRT_h1} and~\eqref{SE_PZF_MRT_l1}.
Finally, to show the benefit of using the hybrid OTFS/OFDM system, we compare the performance of FZF and PZF with OTFS/OFDM to that with OFDM, respectively. The results show that there is a performance improvement for the hybrid OTFS/OFDM system over the OFDM system.
\begin{figure}
\centering
\includegraphics[width=0.43\textwidth]{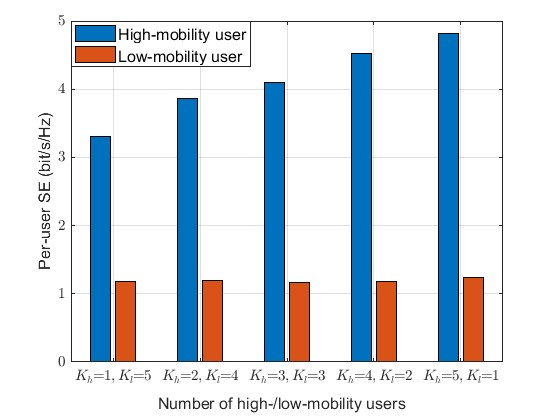}
\caption{Per-user SEs for PZF precoding and HL user grouping and for different numbers of users per each group ($K=6$).}
\vspace{0.5em}
\label{Nrs}
\centering
\end{figure}

\begin{figure*}
{\small
\begin{align}
\label{Dkkp_1}
\mathbb E \left\{ \D_{k_h k_l'} \D_{k_h k_l'}^{\rm H}  \right\}
&={E_s} \mathbb E \left\{{\Bar{\bf{H}}}^{\rm DD}_{k_h} {\bf W}_{k'_l}^\MRT ({\bf W}_{k'_l}^\MRT)^{\rm H} ({\Bar{\bf{H}}}^{\rm DD}_{k_h})^{\rm H}\right\} \notag\\
&={E_s} \alpha_\MRT ^2 \mathbb E \left\{{\Bar{\bf{H}}}^{\rm DD}_{k_h} \left( {{\bf I}_{N_t} \otimes{{\bf{I}}_N} \otimes {{\bf{F}}_M}} \right) (\Htd _{k_l'})^{\rm H}         ({ {{\bf{I}}_N} \otimes ( {\bf{F}}_M^{\rm H}{{\bf{F}}_M} })) \Htd _{k_l'} \left( {{\bf I}_{N_t} \otimes{{\bf{I}}_N} \otimes {{\bf{F}}^{\rm H}_M}} \right) ({\Bar{\bf{H}}}^{\rm DD}_{k_h})^{\rm H}\right\}\notag\\
&\stackrel{(a)}{=}{E_s} \alpha_\MRT ^2({{{\bf{F}}_N} \!\otimes \!{{\bf{I}}_M}})
\sum_{i=1}^{P} \sum_{j=1}^{P} \!\mathbb E \left\{ {\bm \theta} _{k_h(i)} {\bm \theta} _{k_l'(j)}^{\rm H} {\bm \theta} _{k_l'(j)} {\bm \theta} ^{\rm H}_{k_h(i)}\right\}\otimes \mathbb E \left\{ {\bf H}^{\rm TD} _{k_h(i)}  ({\bf H}^{\rm TD} _{k'_l(j)})^{\rm H} {\bf H}^{\rm TD} _{k'_l(j)}  ({\bf H}^{\rm TD} _{k_h(i)})^{\rm H} \right\}  ({{{\bf{F}}_N^{\rm H}} \otimes {{\bf{I}}_M}})  \notag\\
&\stackrel{(b)}=\frac{{E_s} \alpha_\MRT ^2}{P^2}\sum_{i=1}^{P} \sum_{j=1}^{P} 
\mathbb E \left\{ | {\bm \theta} _{k_h(i)} {\bm \theta} _{k_l'(j)}^{\rm H} |^2 \right\} {\bf I}_{MN}
\tag{44}
\end{align} 
}
	\hrulefill
	\vspace{-4mm}
\end{figure*}

To better study the trade-off between the complexity and performance of the PZF scheme, in Fig.~\ref{Nrs}, we show the per-user SE for high/low-mobility users with different numbers of high-mobility users, $K_h$, for a given number of users $K=6$, and $K_l=K-K_h$. We can see that with the increase of $K_h$, higher SE can be achieved by the high-mobility users. This is because, with smaller $K_l$, high-mobility users suffer from less inter-group interference. However, the SE for low-mobility users remains the same, as they are affected by both intra-group and inter-group interference.
We recall that for high-mobility users, extra complexity is required to implement OTFS modulation. Moreover, the complexity of PZF is dependent on $K_h$. With larger $K_h$, higher computational complexity is required for high-mobility users. 

\vspace{-4mm}
\section{Conclusion}
We investigated the performance of a DL massive MIMO-OTFS/OFDM system with different precoding designs, in terms of SE. We considered FZF and PZF precoding design under HL and SW user grouping strategies.  We showed that the FZF eliminates all the interference at the cost of high complexity. We investigated the PZF precoding with two different grouping strategies: one based on high and low mobility user grouping, and the other based on strong and weak user grouping. Our simulation results validated our discussions and demonstrated the trade-off between performance and complexity. 
For our future work, the performance with large-scale fading and imperfect channel state information can be investigated.

\appendices
\vspace{-4mm}
\section{Proof of Proposition 1}
\label{APX:theorem:FZF:LM}
By invoking~\eqref{eq:SEk} and~\eqref{FZF}, for the $k_h$-th high-mobility user, we derive $\D_{k_h k'}$ as
{\small
\begin{align}
	\mH _{k_h}\qW_{k'}^\FZF	\!\!\!&=\! \alpha_{\FZF} \mH _{k_h} ({\mH}^{\rm FZF}_{\rm H})^{\rm H} \left (  {\mH}^{\rm FZF}_{\rm H} ({\mH}^{\rm FZF}_{\rm H})^{\rm H} \right )^{\!\!-1}\!\! \big ( \!{\qb_{K}^{(k')}} \otimes \qI_{MN} \!\big)  \notag \\
	&=  \alpha_{\FZF} \left ( ({{\bf{b}}_{K}^{(k_h)}})^{\rm H} \otimes {\bf I}_{MN} \right)  {\mH}^{\rm FZF}_{\rm H} ({\mH}^{\rm FZF}_{\rm H})^{\rm H}  \notag\\
    &\hspace{5em}\times \left (  {\mH}^{\rm FZF}_{\rm H} ({\mH}^{\rm FZF}_{\rm H})^{\rm H} \right )^{\!\!-1}\big ( {{\bf{b}}_{K}^{(k')}} \otimes {\bf I}_{MN} \big)  \notag \\
	&=\alpha_{\FZF} \Big ( \big (({{\bf{b}}_{K}^{(k_h)}})^{\rm H} {{\bf{b}}_{K}^{(k')}}\big ) \otimes {\bf I}_{MN} \Big)  \notag \\
	&=
	    \begin{cases}
			{\bf 0}_{MN}, & k'\neq k_h\\
			\alpha_{\FZF}{\bf I}_{MN}, & k'= k_h
		\end{cases}.
	\label{FZF_HW_H}
\end{align}
}
Therefore, 
\vspace{-0.3em}
\begin{align}~\label{eq:Dkhkh}
	\BD_{k_h k_h}=\alpha_{\FZF} \sqrt{{E_s}} {\bf I}_{MN},
\end{align}
\begin{align}
\mathbb E \big\{ \D_{k k} \D_{k k}^{\rm H}  \big\}=\alpha_{\FZF}^2 {E_s} {\bf I}_{MN},
\end{align}
and
\begin{align}~\label{eq:SiFZF}
	\Psi_{k_h} \!\!\!=\! {\bf I}_{MN} \!+\! \mathbb E \big\{ \D_{k k} \D_{k k}^{\rm H}  \big\} \!+\!(K-1){\bf 0}_{MN}\!-\! \BD_{k k} \BDH_{k k}
 = {\bf I}_{MN}.
\end{align}
Hence, by substituting~\eqref{eq:Dkhkh} and~\eqref{eq:SiFZF} into~\eqref{eq:SEk}, the SE for $k_h$-th high-mobility user can be obtained as~\eqref{SE_HFZF}. Similarly, for the $k_l$-th low-mobility user, we have 
{\small
 \begin{align}
	\mH _{k_l}\qW_{k'}^\FZF	\!\!\!&=\! \alpha_{\FZF} \mH _{k_l} ({\mH}^{\rm FZF}_{\rm H})^{\rm H} \left (  {\mH}^{\rm FZF}_{\rm H} ({\mH}^{\rm FZF}_{\rm H})^{\rm H} \right )^{\!\!-1}\!\! \left ( \!{\qb_{K}^{(k')}} \otimes \qI_{MN} \!\right)  \notag \\
	&=
	    \begin{cases}
			{\bf 0}_{MN}, & k'\neq k_l\\
			\alpha_{\FZF}{\bf I}_{MN}, & k'= k_l
		\end{cases}.
  \label{FZF_HW_L}
  \end{align}}
Then, the SE for the $k_l$-th low-mobility user can be obtained as~\eqref{SE_LFZF}.
\vspace{-1em}
\section{Proof of Proposition 2}
\label{Ptheorem:PZFMRT}
Focusing on the $k_h$-th high-mobility user, similar as in~\eqref{FZF_HW_H} and~\eqref{FZF_HW_L}, we have
{\small
 \begin{align}
	\Hdd _{k_h}\qW_{k'_h}^\PZF	\!\!\!&=\! \alpha_{\PZF} \Hdd _{k_h} (\mH^\PZF)^{\rm H} \left (  \mH^\PZF (\mH^\PZF)^{\rm H} \right )^{\!-1} \left ( \!{\qb_{K_h}^{(k'_h)}} \otimes \qI_{MN} \!\right)  \notag \\
	&=
	    \begin{cases}
			{\bf 0}_{MN}, & k'_h\neq k_h, k'_h\in \Kh\\
			\alpha_{\PZF}{\bf I}_{MN}, & k'_h= k_h
		\end{cases}.
  \end{align} }
Therefore,
\begin{align}
\BD_{k_h k_h} &=  \sqrt{E_s} \mathbb{E}\left\{  {\qH}^{\rm DD}_{k_h}   {\bf W}_{k_h}^\PZF \right\} =\sqrt{E_s}\alpha_\PZF {\bf I}_{MN}.
\end{align}
Then, for the intra-group interference from user $k_h'$, where $k_h'\in \Kh$, $k_h'\neq k_h$, we have
{\small
\begin{align}
\mathbb E \big\{ \D_{k_h k_h'} \D_{k_h k_h'}^{\rm H}  \big\}&=\mathbb E \big\{  {E_s}  {\qH}^{\rm DD}_{k_h} {\bf W}_{k'_h}^\PZF ({\bf W}_{k'_h}^\PZF)^{\rm H} ({\qH}^{\rm DD}_{k_h})^{\rm H}\big\}={\bf 0}_{MN}.
\end{align}  }
Moreover, for the inter-group interference from user $k_l'$, with $k_l\in \Kl$, $\mathbb E \big\{ \D_{k_h k_l'} \D_{k_h k_l'}^{\rm H}  \big\}$ can be derived as in~\eqref{Dkkp_1} at the top of the page.
Note that $(a)$ in~\eqref{Dkkp_1} is based on~\eqref{MIMOH} and the properties of Kronecker product, while $(b)$ in~\eqref{Dkkp_1} is based on the independence of the channel responses of different paths.

\bibliographystyle{IEEEtran}
\bibliography{Main.bib}

\end{document}